\documentclass[12pt]{article}

\pdfoutput=1

\usepackage[pdftex]{graphicx}
\usepackage{amsmath,amssymb}
\usepackage{psfrag}
\usepackage[pdftex]{hyperref}

\newcounter{comment}

{\refstepcounter{comment}%
\begin{quote}
\ttfamily\small$\blacksquare$ \textbf{\underline{Comment} $\sharp$\thecomment:}}%
{\end{quote}}

{
\begin{quote}
\ttfamily\small$\blacktriangleright$ \textbf{\underline{Reply} $\sharp$\thecomment:}}%
{\end{quote}}

\font\cmss=cmss12 
\def\1{\hbox{{1}\kern-.25em\hbox{l}}}
\def\bfZ{\relax{\hbox{\cmss Z\kern-.4em Z}}}

\def\ru1{\rule[-0.4truecm]{0mm}{1truecm}}

\begin{document}



\setcounter{footnote}{0}
\renewcommand{\thefootnote}{\fnsymbol{footnote}}
\renewcommand{\bar}[1]{\overline{#1}}
\newcommand{\ie}{{\it i.e.}}
\newcommand{\eg}{{\it e.g.,}}
\newcommand{\btt}[1]{{\tt$\backslash$#1}}
\newcommand{\half}{{$\frac{1}{2}$}} 
\newcommand{\ket}[1]{\left\vert\,{#1}\right\rangle}
\newcommand{\VEV}[1]{\left\langle{#1}\right\rangle}

\def\ru1{\rule[-0.4truecm]{0mm}{1truecm}}
\def\upleftarrow#1{\overleftarrow{#1}}
\def\uprightarrow#1{\overrightarrow{#1}}
\def\thru#1{\mathrel{\mathop{#1\!\!\!/}}}

\def\senk#1{\bbox{#1}_\perp}
\def\ha{{1\over 2}}
\def\ub#1{\underline{#1}}
\def\ths{\thinspace}
\def\psibar{\overline{\psi}}
\def\del{\partial}
\def\ra{\rightarrow}
\def\eg{{\it e.g.}}
\def\g{\gamma}

\newpage
\begin{flushright}
October  2007
\end{flushright}

\bigskip

{\centerline{\Large \bf A Method of Experimentally Probing}
{\centerline{\Large \bf Transeverse Momentum Dependent Distributions}}


\vspace{7mm}
\vspace{7mm}

\centerline{\bf Dae Sung Hwang$^{a}$ and Dong Soo Kim$^{b}$}

\vspace{3mm}

\vspace{4mm} \centerline{\it $^a$Department of Physics, Sejong University, Seoul
143--747, South Korea}
\vspace{1mm} \centerline{\it $^b$Department of Physics, Kangnung National University,
Kangnung 210-702, South Korea}

\vspace*{1.2cm}



\begin{abstract}
\noindent
We calculate the double spin asymmetry $A_{LL}(x, y, z, P_{hT})$ of $\pi^0$ production
with the spectator model and the model based on the factorization ansatz.
We also calculate the double spin asymmetry for the integration over the
range of $(x,y,z)$ for the setups of the experiments of COMPASS, HERMES, and JLab.
We find that the results are characteristically dependent on the model used.
Therefore, we suggest that the measurements of the double spin asymmetry provides
a method of experimentally probing the transeverse momentum dependent distributions.
\end{abstract}

\vfill

\centerline{
PACS numbers:
12.39.Ki, 13.60.-r, 13.60.Le, 13.88.+e
}
\vfill


\newpage

\section{Introduction}
In recent years the role of the transverse momentum of the parton has
been more important in the field of the hadron physics since, for example,
it provides time-odd distribution and fragmentation functions,
and makes the single-spin asymmetries in hadronic processes possible
\cite{MT96,BHS}.
The transverse momentum of the parton inside the proton is also related to
the orbital angular momentum carried by the parton, which is an important
subject since it is considered as a part of the spin contents of the proton.
It is important to probe experimentally how the distribution and fragmentation
functions are dependent on the transverse momenta of partons.


In a lot of researches on the transverse momentum dependent
distribution(fragmentation) functions, the ansatz which factorizes $x$($z$) and
$k_{\perp}$($p_{\perp}$) is adopted.
For example, Ref. \cite{anselmino06} investigated the double spin asymmetry $A_{LL}$
by using such factorized distribution and fragmentation functions.
We study the differences of the distribution and fragmentation functions of
the factorized model and those of the spectator model.

Jakob et al. \cite{mulders97} presented a spectator model,
which is based on the scalar and axial-vector diquark models
of the nucleon.
The important character of the spectator model is that
the longitudinal momentum fraction $x$ and the transverse momentum $k_{\perp}$
of the parton
are intimately correlated with each other, since the spectator model is based on
Lorentz invariant Feynman diagram.
The transverse momentum distributions of the up and down
quarks inside the proton are different, since for the proton
the up quark is composed of a linear combination of
the scalar and axial-vector diquark components and the down quark is only
composed of the axial-vector diquark component.

We calculate the dependence of the double spin asymmetry $A_{LL}$ of $\pi^0$ production
on the variables $x$, $y$, $z$ and $P_{hT}$ with the spectator model,
and find that the results are characteristically different from those caculated
with the model based on the factorization ansatz.
For example, the $P_{hT}$-behavior of $A_{LL}$ is not sensitive to $z$-value
in the case of the spectator model, whereas it is very sensitive to $z$-value
in the case of the model based on the factorization ansatz.
We also calculate $A_{LL}$ of $\pi^0$ production for the integration over the
range of $(x,y,z)$ for the setups of the experiments of COMPASS, HERMES, and JLab.
The $P_{hT}$-behavior of these results of $A_{LL}$ are also different for the
two models.
Therefore, it should be possible to use such differences in order to discriminate
experimentally the spectator model and the model based on the factorization ansatz.
Then, we suggest that we can discriminate experimentally these two models by
measuring $A_{LL}(x, y, z, P_{hT})$ to obtain the information on
which model is closer to the physical reality.

In section 2 we present the results of $A_{LL}$ obtained by calculating with
the spectator model of Jakob et al.
In section 3 we calculate $A_{LL}$ by using the model based on the factorization ansatz,
and compare the results with those of section 2.
Section 4 is conclusion.

\section{Spectator Model}
\subsection{Distribution and Fragmentation Functions}
Jakob et al. \cite{mulders97} presented a spectator model,
which is based on the scalar and axial-vector diquark models
of the nucleon.
The important character of the spectator model is that
the longitudinal momentum fraction $x$ and the transverse momentum $k_{\perp}$
of the parton
are intimately correlated with each other, since the spectator model is based on
Lorentz invariant Feynman diagram.
In this model the unpolarized and polarized distribution functions
$f_1$ and $g_1$ are given by
\begin{equation}
f_{1R}(x, {\bf k}_{\perp})=
N_R\
{(xM+m)^2+{\bf k}_{\perp}^2\over ({\bf k}_{\perp}^2+{\lambda}_R^2)^{2\alpha}}\ ,
\ \ \
g_{1R}(x, {\bf k}_{\perp})=N_R\
a_R\
{(xM+m)^2-{\bf k}_{\perp}^2\over ({\bf k}_{\perp}^2+{\lambda}_R^2)^{2\alpha}}\ ,
\label{f1g1jmr}
\end{equation}
where
${\lambda}_R^2(x)=(1-x){\Lambda}^2+xM_R^2-x(1-x)M^2$ and
$a_s=1$, $a_a=-{1\over 3}$ for $a_R$,
with $M=0.94$ GeV,
$M_s=0.6$ GeV, $M_a=0.8$ GeV for $M_R$.
We take $\alpha = 2$ in this paper.
In this subsection we take $\Lambda =0.5$ GeV. However, in the next subsection
we also consider 0.4 and 0.6 GeV for the $\Lambda$ value.
Here the subscripts $s$ and $a$ refer to the scalar and axial-vector diquarks
The normalization constant $N_R$ is fixed by the normalization condition of
$f_{1R}(x, {\bf k}_{\perp})$.

From the $SU(4)$ wave function of the proton, we have (also for $g_1^q$)
\cite{mulders97}
\begin{equation}
f_1^u={3\over 2}f_{1s}+{1\over 2}f_{1a}\ ,
\ \ \
f_1^d=f_{1a}\ .
\label{f1udab}
\end{equation}
That is, transverse momentum distributions of the up and down
quarks inside the proton are different, since for the proton
the up quark is composed of a linear combination of
the scalar and axial-vector diquark components and the down quark is only
composed of the axial-vector diquark component.
Then, Eqs. (\ref{f1g1jmr}) and (\ref{f1udab}) give
\begin{eqnarray}
f_1^u(x, {\bf k}_{\perp})&=&
{3\over 2}N_s{(xM+m)^2+{\bf k}_{\perp}^2\over ({\bf k}_{\perp}^2+{\lambda}_s^2)^{2\alpha}}
+{1\over 2}N_a{(xM+m)^2+{\bf k}_{\perp}^2\over ({\bf k}_{\perp}^2+{\lambda}_a^2)^{2\alpha}},
\label{fgu1}\\
g_1^u(x, {\bf k}_{\perp})&=&
{3\over 2}N_s{(xM+m)^2-{\bf k}_{\perp}^2\over ({\bf k}_{\perp}^2+{\lambda}_s^2)^{2\alpha}}
-{1\over 6}N_a{(xM+m)^2-{\bf k}_{\perp}^2\over ({\bf k}_{\perp}^2+{\lambda}_a^2)^{2\alpha}},
\nonumber\\
f_1^d(x, {\bf k}_{\perp})&=&
N_a{(xM+m)^2+{\bf k}_{\perp}^2\over ({\bf k}_{\perp}^2+{\lambda}_a^2)^{2\alpha}},
\nonumber\\
g_1^d(x, {\bf k}_{\perp})&=&
-{1\over 3}N_a{(xM+m)^2-{\bf k}_{\perp}^2\over ({\bf k}_{\perp}^2+{\lambda}_a^2)^{2\alpha}}.
\nonumber
\end{eqnarray}
The distribution functions given in (\ref{fgu1}) are plotted in Figs. 1 and 2.
Fig. 3 presents the widths of the distribution functions in ${\bf k}_{\perp}$
as functions of $x$.

We use for both $u$ and $d$ quarks the fragmentation function given in Ref. \cite{abm05},
which is plotted in Fig. 4:
\begin{equation}
D_1(z,{\bf p}_{\perp})={1\over z}{g^2\over 16\pi^3}
{{\bf p}_{\perp}^2+m^2\over ({\bf p}_{\perp}^2+m^2+{1-z\over z^2}m_{\pi}^2)^2}\ ,
\label{d1ff}
\end{equation}
where $m_\pi$ is pion mass and $m=0.3$ GeV.

\begin{figure}
\centering
\psfrag{xperp}[cc][cc]{$x_\perp$}
\begin{minipage}[t]{6.0cm}
\centering
\includegraphics[width=\textwidth]{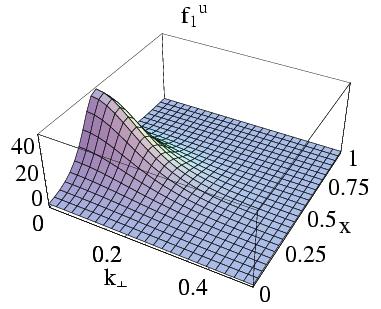}
(a)
\end{minipage}\hspace{1.0cm}
\begin{minipage}[t]{6.0cm}
\centering
\includegraphics[width=\textwidth]{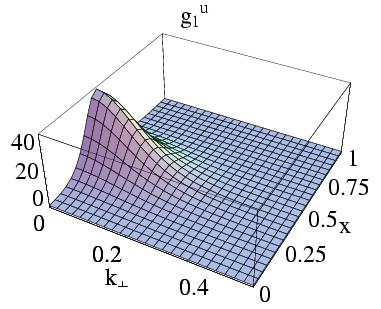}
(b)
\end{minipage}
\parbox{0.95\textwidth}{\caption{
The distribution functions of $u$ quark
$f_1^u(x, {\bf k}_{\perp})$ (left) and $g_1^u(x, {\bf k}_{\perp})$ (right)
given in (\ref{fgu1}).
\label{comparison1}}}
\end{figure}

\begin{figure}
\centering
\psfrag{xperp}[cc][cc]{$x_\perp$}
\begin{minipage}[t]{6.0cm}
\centering
\includegraphics[width=\textwidth]{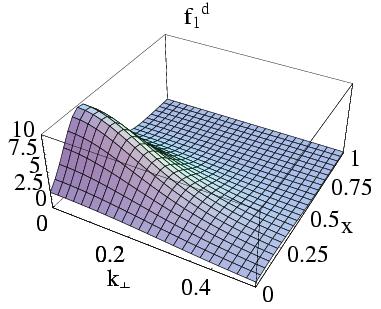}
(a)
\end{minipage}\hspace{1.0cm}
\begin{minipage}[t]{6.0cm}
\centering
\includegraphics[width=\textwidth]{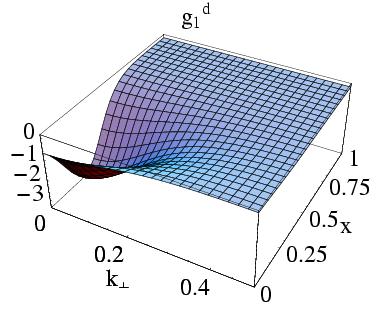}
(b)
\end{minipage}
\parbox{0.95\textwidth}{\caption{
The distribution functions of $d$ quark
$f_1^d(x, {\bf k}_{\perp})$ (left) and $g_1^d(x, {\bf k}_{\perp})$ (right)
given in (\ref{fgu1}).
\label{comparison2}}}
\end{figure}

\begin{figure}
\centering
\psfrag{xperp}[cc][cc]{$x_\perp$}
\begin{minipage}[t]{6.0cm}
\centering
\includegraphics[width=\textwidth]{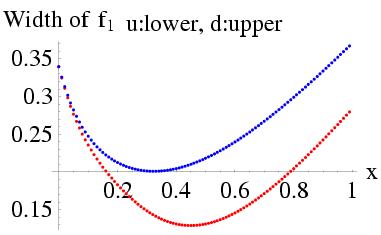}
(a)
\end{minipage}\hspace{1.0cm}
\begin{minipage}[t]{6.0cm}
\centering
\includegraphics[width=\textwidth]{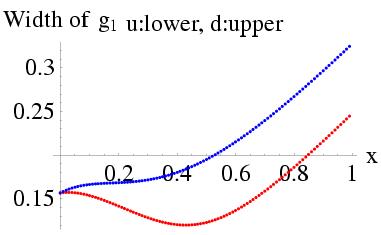}
(b)
\end{minipage}
\parbox{0.95\textwidth}{\caption{
The width in ${\bf k}_{\perp}$, which is defined as the value
of ${\bf k}_{\perp}$ which satisfies $f_1(g_1)(x, {\bf k}_{\perp})=
{1\over 2}f_1(g_1)(x, {\bf 0}_{\perp})$ for $u$ quark (lower line) and
$d$ quark (upper line).
\label{comparison3}}}
\end{figure}

\begin{figure}
\centering
\psfrag{xperp}[cc][cc]{$x_\perp$}
\begin{minipage}[t]{6.0cm}
\centering
\includegraphics[width=\textwidth]{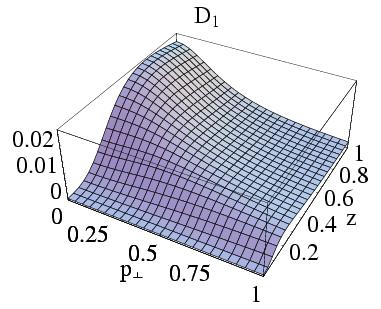}
\end{minipage}
\parbox{0.95\textwidth}{\caption{
The fragmentation function
$D_1(z,{\bf p}_{\perp})$ given in (\ref{d1ff}).
\label{comparison4}}}
\end{figure}

\begin{figure}
\centering
\psfrag{xperp}[cc][cc]{$x_\perp$}
\begin{minipage}[t]{5.7cm}
\centering
\includegraphics[width=\textwidth]{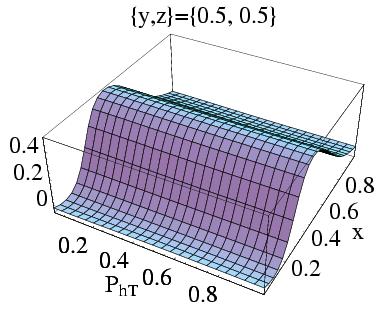}
(a1)
\end{minipage}\hspace{0.0cm}
\begin{minipage}[t]{5.7cm}
\centering
\includegraphics[width=\textwidth]{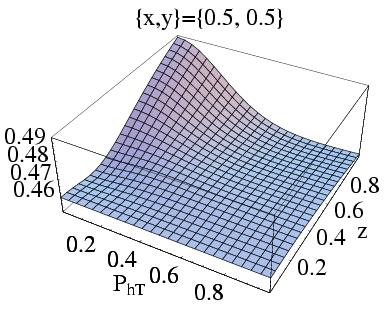}
(b1)
\end{minipage}\hspace{0.0cm}
\begin{minipage}[t]{5.7cm}
\centering
\includegraphics[width=\textwidth]{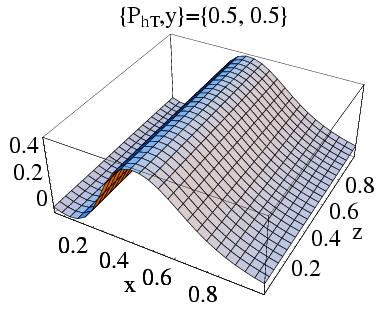}
(c1)
\end{minipage}
\begin{minipage}[t]{5.7cm}
\centering
\includegraphics[width=\textwidth]{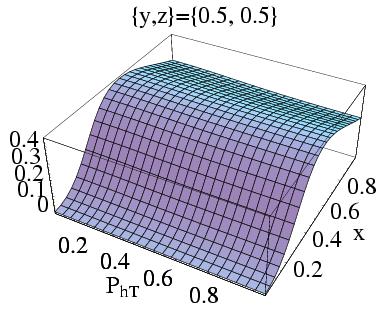}
(a2)
\end{minipage}\hspace{0.0cm}
\begin{minipage}[t]{5.7cm}
\centering
\includegraphics[width=\textwidth]{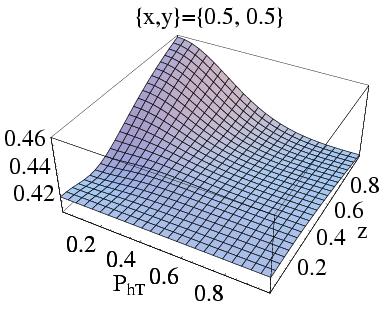}
(b2)
\end{minipage}\hspace{0.0cm}
\begin{minipage}[t]{5.7cm}
\centering
\includegraphics[width=\textwidth]{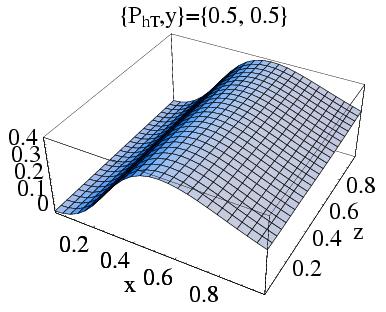}
(c2)
\end{minipage}
\begin{minipage}[t]{5.7cm}
\centering
\includegraphics[width=\textwidth]{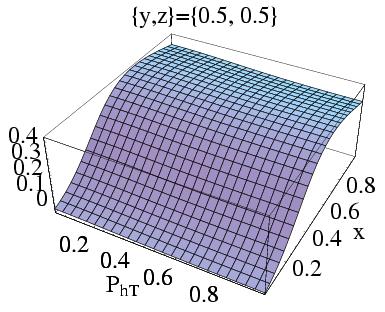}
(a3)
\end{minipage}\hspace{0.0cm}
\begin{minipage}[t]{5.7cm}
\centering
\includegraphics[width=\textwidth]{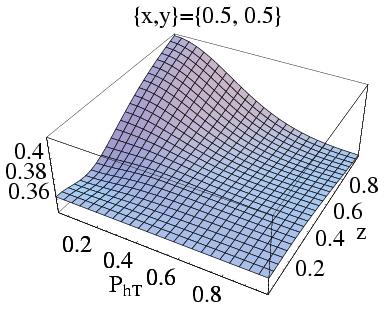}
(b3)
\end{minipage}\hspace{0.0cm}
\begin{minipage}[t]{5.7cm}
\centering
\includegraphics[width=\textwidth]{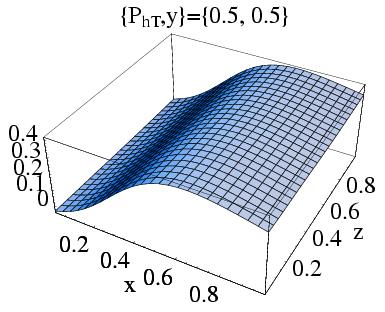}
(c3)
\end{minipage}
\parbox{0.95\textwidth}{\caption{
For the spectator model, $A_{LL}$ of $\pi^0$ production as a function of
$P_{hT}$ and $x$ (a) with fixed $y=0.5$ and $z=0.5$,
that of $P_{hT}$ and $z$ (b) with fixed $x=0.5$ and $y=0.5$,
and that of $x$ and $z$ (c) with fixed $P_{hT}=0.5$ and $y=0.5$.
$\Lambda=0.4$ GeV (a1, b1, c1),
$\Lambda=0.5$ GeV (a2, b2, c2),
and $\Lambda=0.6$ GeV (a3, b3, c3).
\label{comparison5}}}
\end{figure}

\begin{figure}
\centering
\psfrag{xperp}[cc][cc]{$x_\perp$}
\begin{minipage}[t]{5.7cm}
\centering
\includegraphics[width=\textwidth]{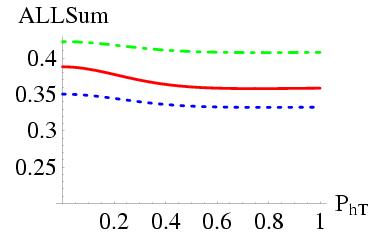}
(1)
\end{minipage}\hspace{0.0cm}
\begin{minipage}[t]{5.7cm}
\centering
\includegraphics[width=\textwidth]{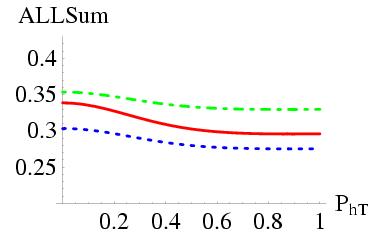}
(2)
\end{minipage}\hspace{0.0cm}
\begin{minipage}[t]{5.7cm}
\centering
\includegraphics[width=\textwidth]{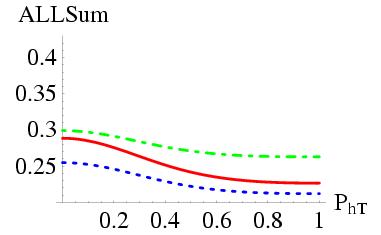}
(3)
\end{minipage}
\parbox{0.95\textwidth}{\caption{
For the spectator model,
$A_{LL}$ of ${\pi}^0$ production
for the integration over the ranges of $(x,y,z)$ for the
setups of the experiments of COMPASS(solid), HERMES(dotted),
and JLab(dash-dotted line).
(1) for $\Lambda =0.4$ GeV;
(2) for $\Lambda =0.5$ GeV;
(3) for $\Lambda =0.6$ GeV.
\label{comparison6}}}
\end{figure}

\begin{figure}
\centering
\psfrag{xperp}[cc][cc]{$x_\perp$}
\begin{minipage}[t]{5.7cm}
\centering
\includegraphics[width=\textwidth]{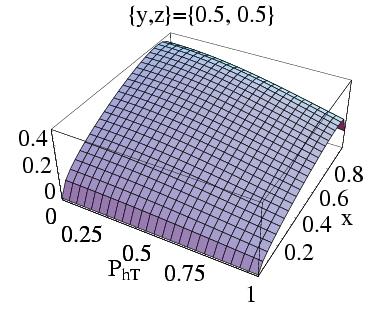}
(a1)
\end{minipage}\hspace{0.0cm}
\begin{minipage}[t]{5.7cm}
\centering
\includegraphics[width=\textwidth]{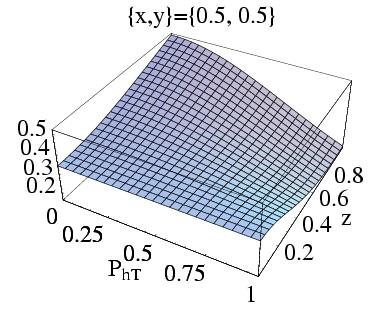}
(b1)
\end{minipage}\hspace{0.0cm}
\begin{minipage}[t]{5.7cm}
\centering
\includegraphics[width=\textwidth]{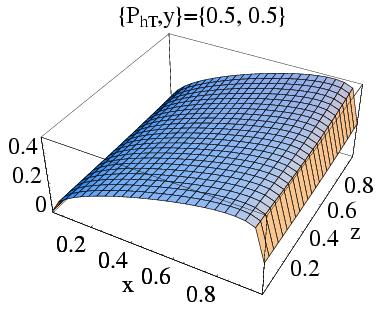}
(c1)
\end{minipage}
\begin{minipage}[t]{5.7cm}
\centering
\includegraphics[width=\textwidth]{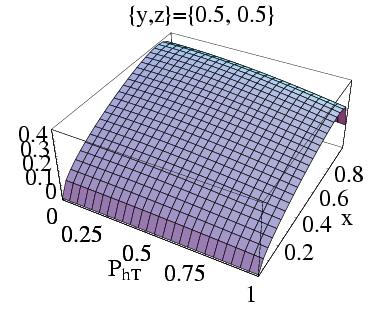}
(a2)
\end{minipage}\hspace{0.0cm}
\begin{minipage}[t]{5.7cm}
\centering
\includegraphics[width=\textwidth]{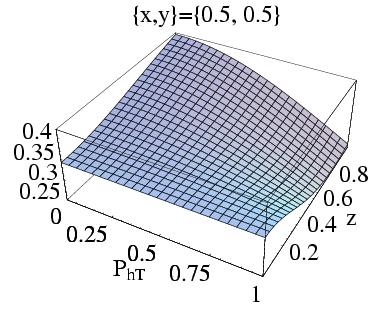}
(b2)
\end{minipage}\hspace{0.0cm}
\begin{minipage}[t]{5.7cm}
\centering
\includegraphics[width=\textwidth]{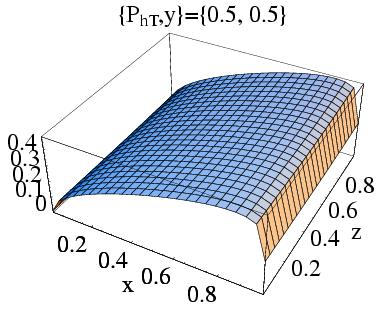}
(c2)
\end{minipage}
\begin{minipage}[t]{5.7cm}
\centering
\includegraphics[width=\textwidth]{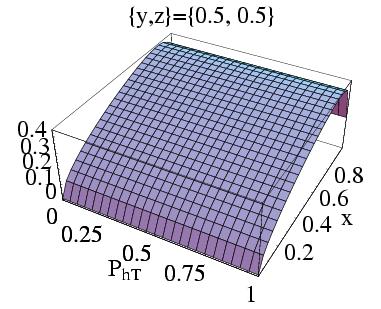}
(a3)
\end{minipage}\hspace{0.0cm}
\begin{minipage}[t]{5.7cm}
\centering
\includegraphics[width=\textwidth]{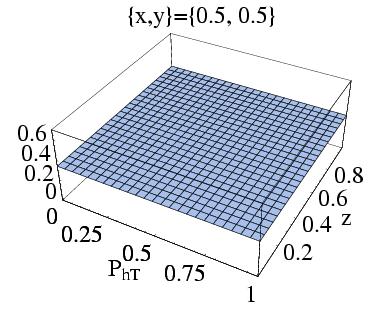}
(b3)
\end{minipage}\hspace{0.0cm}
\begin{minipage}[t]{5.7cm}
\centering
\includegraphics[width=\textwidth]{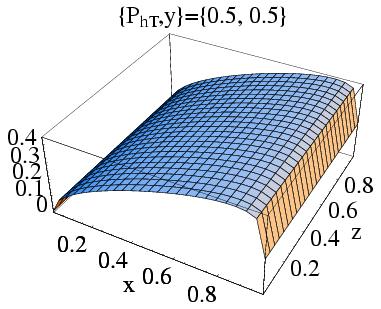}
(c3)
\end{minipage}
\parbox{0.95\textwidth}{\caption{
For the model based on the factorization Ansatz,
$A_{LL}$ of $\pi^0$ production as a function of
$P_{hT}$ and $x$ (a) with fixed $y=0.5$ and $z=0.5$,
that of $P_{hT}$ and $z$ (b) with fixed $x=0.5$ and $y=0.5$,
and that of $x$ and $z$ (c) with fixed $P_{hT}=0.5$ and $y=0.5$.
(a1, b1, c1) for $\mu_2^2 =0.10$ GeV${}^2$;
(a2, b2, c2) for $\mu_2^2 =0.17$ GeV${}^2$;
(a3, b3, c3) for $\mu_2^2 =0.25$ GeV${}^2$.
\label{comparison5}}}
\end{figure}

\begin{figure}
\centering
\psfrag{xperp}[cc][cc]{$x_\perp$}
\begin{minipage}[t]{5.7cm}
\centering
\includegraphics[width=\textwidth]{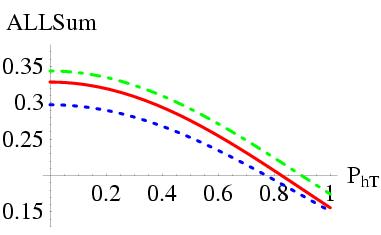}
(1)
\end{minipage}\hspace{0.0cm}
\begin{minipage}[t]{5.7cm}
\centering
\includegraphics[width=\textwidth]{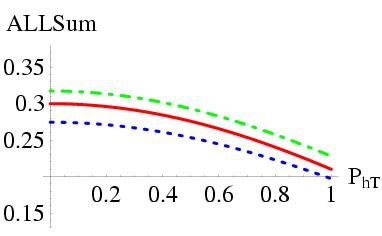}
(2)
\end{minipage}\hspace{0.0cm}
\begin{minipage}[t]{5.7cm}
\centering
\includegraphics[width=\textwidth]{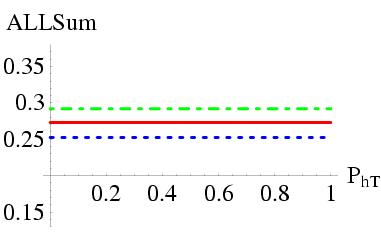}
(3)
\end{minipage}
\parbox{0.95\textwidth}{\caption{
For the model based on the factorization Ansatz,
$A_{LL}$ of ${\pi}^0$ production
for the integration over the ranges of $(x,y,z)$ for the
setups of the experiments of COMPASS(solid), HERMES(dotted),
and JLab(dash-dotted line).
(1) for $\mu_2^2 =0.10$ GeV${}^2$;
(2) for $\mu_2^2 =0.17$ GeV${}^2$;
(3) for $\mu_2^2 =0.25$ GeV${}^2$.
\label{comparison6a}}}
\end{figure}

\subsection{Double Spin Asymmetry}
The double spin asymmetry is given by \cite{anselmino06}
\begin{equation}
A_{LL}(x,y,z,P_{hT})={\Delta \sigma_{LL}\over \sigma_0}\ ,
\label{a1}
\end{equation}
where
\begin{eqnarray}
\Delta\sigma_{LL}&=&{\pi\over xy^2}\ \Big[\ y(2-y)\ \Big]\
\Sigma_q e_q^2\ \int d^2{\bf k}_{\perp}\ g_{1}^q(x,{\bf k}_{\perp})\
D_q^h(z,{\bf P}_{hT}-z{\bf k}_{\perp})
\ ,
\label{a2}\\
\sigma_0&=&{\pi\over xy^2}\ \Big[\ 1+(1-y)^2\ \Big]\
\Sigma_q e_q^2\ \int d^2{\bf k}_{\perp}\ f_{1}^q(x,{\bf k}_{\perp})\
D_q^h(z,{\bf P}_{hT}-z{\bf k}_{\perp})
\ .
\nonumber
\end{eqnarray}

We study $A_{LL}(x,y,z,P_{hT})$ with the model of Ref. \cite{mulders97}
by using
$f_{1}^q(x,{\bf k}_{\perp})$ and $g_{1}^q(x,{\bf k}_{\perp})$
given in (\ref{fgu1}), and $D_q^h(z,{\bf p}_{\perp})$
given in (\ref{d1ff}) for both $u$ and $d$ quarks.
We consider only the contributions from the valence quarks $u$ and $d$,
and ignore the contributions from sea quarks.
We calculate the dependence of the double spin asymmetry $A_{LL}(x,y,z,P_{hT})$
of $\pi^0$ production
on the variables $x$, $y$, $z$ and $P_{hT}$ with the spectator model.
We consider three values 0.4, 0.5, 0.6 GeV for $\Lambda$ existing in (\ref{fgu1})
through $\lambda_R$,
in order to see the sensitivity of the results to the parameter value of $\Lambda$.
The results of the calculation are presented in Fig. 5.

We also calculate $A_{LL}$ of $\pi^0$ production for the integration over the
range of $(x,y,z)$ for the setups of the experiments of COMPASS, HERMES, and JLab.
The following ranges are covered by the setup of each experiment \cite{anselmino06},\\
(A) COMPASS: $0.1 < x < 0.6$, $0.5 < y < 0.9$, and $0.4 < z < 0.9$;\\
(B) HERMES:  $0.1 < x < 0.6$, $0.45 < y < 0.85$, and $0.4 < z < 0.7$;\\
(C) JLab:    $0.2 < x < 0.6$, $0.4 < y < 0.85$, and $0.4 < z < 0.7$.\\
The results are presented in Fig. 6 for three values of $\Lambda$:
0.4, 0.5, 0.6 GeV.

\section{Comparison With Model Based On Factorization}
\subsection{Distribution and Fragmentation Functions}
For the transverse momentum dependent distribution(fragmentation) functions,
the ansatz which factorizes $x$($z$) and $k_{\perp}$($p_{\perp}$) is often adopted. 
For example, Ref. \cite{anselmino06} 
used the factorized functions given by
\begin{eqnarray}
f_1^q(x,{\bf k}_{\perp})&=&
f_1^q(x){1\over \pi{\mu}_0^2}{\rm exp}\Big( -{{\bf k}_{\perp}^2\over {\mu}_0^2}\Big)
\ ,
\label{a1factor}\\
g_1^q(x,{\bf k}_{\perp})&=&
g_1^q(x){1\over \pi{\mu}_2^2}{\rm exp}\Big( -{{\bf k}_{\perp}^2\over {\mu}_2^2}\Big)
\ ,
\nonumber\\
D_q^h(z,{\bf p}_{\perp})&=&
D_q^h(z){1\over \pi{\mu}_D^2}{\rm exp}\Big( -{{\bf p}_{\perp}^2\over {\mu}_D^2}\Big)
\ .
\nonumber
\end{eqnarray}
If we draw graphs for the width in ${\bf k}_{\perp}$ corresponding to Fig. 3
in the case of (\ref{a1factor}),
we would get graphs of constants.

For the integrated parton distribution functions appearing in (\ref{a1factor}),
we use the following functions \cite{glueck98,glueck01},
\begin{eqnarray}
xf_1^u(x)&=&xu_v(x,\mu_{\rm NLO}^2)=
0.632 x^{0.43}(1-x)^{3.09}(1+18.2x)
\label{glueck1}\\
xf_1^d(x)&=&xd_v(x,\mu_{\rm NLO}^2)=
0.624(1-x)^{1.0}xu_v(x,\mu_{\rm NLO}^2)
\nonumber\\
g_1^u(x)&=&\delta u(x,{\mu}^2)=
1.019x^{0.52}(1-x)^{0.12}u_v(x,\mu_{\rm NLO}^2)
\nonumber\\
g_1^d(x)&=&\delta d(x,{\mu}^2)=
-0.669x^{0.43}d_v(x,\mu_{\rm NLO}^2) \ .
\nonumber
\end{eqnarray}
For the integrated fragmentation function appearing in (\ref{a1factor}),
we use the following function \cite{kretzer00},
\begin{equation}
D_q^h(z)=D_{u{\bar{d}}}^{\pi^+}(z,{\mu}_0^2)=
N_u^{\pi}z^{-0.829}(1-z)^{0.949}\ .
\label{kretzer1}
\end{equation}

\subsection{Double Spin Asymmetry}
When one uses the factorized distribution and fragmentation functions
given in (\ref{a1factor}) for the calculation of
$\Delta\sigma_{LL}$ and $\sigma_0$
in (\ref{a2}), one has
\begin{eqnarray}
\Delta\sigma_{LL}&=&
{y(2-y)\over xy^2}{1\over {\mu}_D^2+z^2{\mu}_2^2}{\rm exp}
\Big( -{{\bf P}_{hT}^2\over {\mu}_D^2+z^2{\mu}_2^2}\Big)
\ \Sigma_q e_q^2 g_1^q(x) D_q^h(z)
\ ,
\label{a3factor}\\
\nonumber\\
\sigma_{0}&=&
{1+(1-y)^2\over xy^2}{1\over {\mu}_D^2+z^2{\mu}_0^2}{\rm exp}
\Big( -{{\bf P}_{hT}^2\over {\mu}_D^2+z^2{\mu}_0^2}\Big)
\ \Sigma_q e_q^2 f_1^q(x) D_q^h(z)
\ .
\nonumber
\end{eqnarray}
Then, using $\Delta\sigma_{LL}$ and $\sigma_0$ in (\ref{a3factor}),
one can calculate the double spin asymmetry $A_{LL}(x,y,z,P_{hT})$ from (\ref{a1}).
Following Ref. \cite{anselmino06}
we use $\mu_0^2=0.25$ ${\rm GeV}^2$, $\mu_D^2=0.20$ ${\rm GeV}^2$,
and three different values: $\mu_2^2=0.10,\ 0.17,\ 0.25$ ${\rm GeV}^2$.
The results are presented in Fig. 7.
We find that the graphs in Fig. 7 are characteristically different from
the graphs in Fig. 5 which were obtained by using the spectator model.
For example, the $P_{hT}$-behavior of $A_{LL}$ is not sensitive to $z$-value
in the case of the spectator model, whereas it is very sensitive to $z$-value
in the case of the model based on the factorization ansatz.

We also calculate $A_{LL}$ of $\pi^0$ production for the integration over the
range of $(x,y,z)$ for the setups of the experiments of COMPASS, HERMES, and JLab.
The results are presented in Fig. 8, which agree with the graphs in FIG. 1
of Ref. \cite{anselmino06}.
The $P_{hT}$-behavior of the integrated $A_{LL}$ presented in
Fig. 6 and Fig. 8 are also different for the two models.
Therefore, it should be possible to use such differences for discriminating
experimentally the spectator model and the model based on the factorization ansatz.
Then, we suggest that we can discriminate experimentally these two models by
measuring $A_{LL}(x, y, z, P_{hT})$ to obtain the information on
which model is closer to the physical reality.

\section{Conclusion}


Recently it is realized that it is important to know the
transverse momentum dependence of the distributions of partons inside the nucleon.
At first it should be useful to know how realistic the factorization Ansatz is.
In this context, it should be useful to be able to discriminate the spectator model
and the model based on the factorization ansatz.
We found that the double spin asymmetries $A_{LL}(x, y, z, P_{hT})$ obtained by
using the two models are characteristically different from each other.
Therefore, we suggest that the measurement of $A_{LL}(x, y, z, P_{hT})$
can be used as an experimental discrimination of the two models.

We note that Ref. \cite{hk07} studied a related subject in the generalized parton
distributions (GPDs). It showed that the GPDs derived from the spectator
model and those from the model
based on factorizing the $t$-dependence of GPDs give different
properties of the form factors and the reaction amplitudes.

\section*{Acknowledgments}
We wish to thank Harut Avagyan and Stan Brodsky for illuminating discussions.
This work was supported in part by the International Cooperation
Program of the KICOS (Korea Foundation for International Cooperation
of Science \& Technology),
and in part by the 2007 research fund from Kangnung National University.

\end{document}